\documentclass[superscriptaddress,aps,preprintnumbers,showpacs,prd,nofootinbib,preprint]{revtex4-1}

\usepackage{graphicx} 
\usepackage{amsmath,amssymb,amsthm,slashed,bm}
\usepackage{xcolor}
\usepackage[colorlinks,allcolors=blue]{hyperref}

\begin{document}


\renewcommand{\figurename}{Fig.}
\renewcommand{\tablename}{Table.}
\newcommand{\Slash}[1]{{\ooalign{\hfil#1\hfil\crcr\raise.167ex\hbox{/}}}}
\newcommand{\bra}[1]{ \langle {#1} | }
\newcommand{\ket}[1]{ | {#1} \rangle }
\newcommand{\bef}{\begin{figure}}  \newcommand{\eef}{\end{figure}}
\newcommand{\bec}{\begin{center}}  \newcommand{\eec}{\end{center}}
\newcommand{\laq}[1]{\label{eq:#1}}  
\newcommand{\dd}[1]{{d \o d{#1}}}
\newcommand{\Eq}[1]{Eq.~(\ref{eq:#1})}
\newcommand{\Eqs}[1]{Eqs.~(\ref{eq:#1})}
\newcommand{\eq}[1]{(\ref{eq:#1})}
\newcommand{\Sec}[1]{Sec.\ref{chap:#1}}
\newcommand{\ab}[1]{\left|{#1}\right|}
\newcommand{\vev}[1]{ \left\langle {#1} \right\rangle }
\newcommand{\bs}[1]{ {\boldsymbol {#1}} }
\newcommand{\lac}[1]{\label{chap:#1}}
\newcommand{\SU}[1]{{\rm SU{#1} } }
\newcommand{\SO}[1]{{\rm SO{#1}} }
\def\({\left(}
\def\){\right)}
\def\dt{{d \o dt}}
\def\diag{\mathop{\rm diag}\nolimits}
\def\Spin{\mathop{\rm Spin}}
\def\O{\mathcal{O}}
\def\U{\mathop{\rm U}}
\def\Sp{\mathop{\rm Sp}}
\def\SL{\mathop{\rm SL}}
\def\tr{\mathop{\rm tr}}
\def\ebq{\end{equation} \begin{equation}}
\newcommand{\OR}{~{\rm or}~}
\newcommand{\AND}{~{\rm and}~}
\newcommand{\EV}{ {\rm \, eV} }
\newcommand{\KEV}{ {\rm \, keV} }
\newcommand{\MEV}{ {\rm \, MeV} }
\newcommand{\GEV}{ {\rm \, GeV} }
\newcommand{\TEV}{ {\rm \, TeV} }
\def\o{\over}
\def\a{\alpha}
\def\b{\beta}
\def\c{\varepsilon}
\def\d{\delta}
\def\e{\epsilon}
\def\f{\phi}
\def\g{\gamma}
\def\h{\theta}
\def\k{\kappa}
\def\l{\lambda}
\def\m{\mu}
\def\n{\nu}
\def\p{\psi}
\def\q{\partial}
\def\r{\rho}
\def\s{\sigma}
\def\t{\tau}
\def\u{\upsilon}
\def\w{\omega}
\def\x{\xi}
\def\y{\eta}
\def\z{\zeta}
\def\D{\Delta}
\def\G{\Gamma}
\def\H{\Theta}
\def\L{\Lambda}
\def\F{\Phi}
\def\P{\Psi}
\def\S{\Sigma}
\def\me{\mathrm e}
\def\ol{\overline}
\def\tl{\tilde}
\def\*{\dagger}

\preprint{TU-1237}

\title{ 
Induced Domain Walls of QCD Axion, and Gravitational Waves
}

\author{
Junseok Lee
}
\affiliation{Department of Physics, Tohoku University, 
Sendai, Miyagi 980-8578, Japan} 
\author{
Kai Murai
}
\affiliation{Department of Physics, Tohoku University, 
Sendai, Miyagi 980-8578, Japan} 
\author{
Fuminobu Takahashi
}
\affiliation{Department of Physics, Tohoku University, 
Sendai, Miyagi 980-8578, Japan} 
\author{
Wen Yin
}
\affiliation{Department of Physics, Tokyo Metropolitan University, Tokyo 192-0397, Japan}
\affiliation{Department of Physics, Tohoku University, 
Sendai, Miyagi 980-8578, Japan}

\begin{abstract}
We show that heavy axion domain walls induce domain walls of the QCD axion through a mixing between the heavy axion and the QCD axion, even when the pre-inflationary initial condition is assumed for the QCD axion.  The induced domain walls arise because the effective $\theta$ parameter changes across the heavy axion domain walls, shifting the potential minimum of the QCD axion. When the heavy axion domain walls collapse, the induced QCD axion domain walls collapse as well. This novel mechanism for producing the QCD axions can explain dark matter even with the axion decay constant as small as ${\cal O}(10^{9})$ GeV. In particular, this scenario requires domain wall collapse near the QCD crossover, potentially accounting for the stochastic gravitational wave background suggested by recent pulsar timing array observations, including NANOGrav. Using this mechanism, it is also possible to easily create induced domain walls for string axions or axions with a large decay constant, which would otherwise be challenging. We also comment on the implications for cosmic birefringence using induced axion domain walls.
\end{abstract}

\maketitle
\flushbottom

\vspace{1cm}

\section{Introduction}
\label{sec: intro}

The QCD axion~\cite{Weinberg:1977ma,Wilczek:1977pj} is a pseudo-Nambu-Goldstone boson arising in the Peccei-Quinn (PQ) mechanism~\cite{Peccei:1977hh,Peccei:1977ur}, which is a solution to the strong CP problem. See Refs.~\cite{Kim:2008hd,Arias:2012az,Marsh:2015xka,DiLuzio:2020wdo,OHare:2024nmr} for reviews.
Beyond its original motivation, the QCD axion is one of the promising candidates for dark matter.
The production mechanism of QCD axion dark matter is classified by the timing of the spontaneous breaking of the PQ symmetry.

If the PQ symmetry is broken before or during inflation,  the axion will have a nearly uniform field value after inflation. This is called the pre-inflationary scenario.   In this case, the QCD axion can be produced via the misalignment mechanism~\cite{Preskill:1982cy,Abbott:1982af,Dine:1982ah}. Once the axion acquires a mass around the QCD crossover, it starts to oscillate coherently, composing dark matter.
The abundance of the QCD axion depends on the initial field value and the decay constant $f_a$.
If $f_a  \sim 10^{12}$\,GeV, the QCD axion can account for all dark matter without fine-tuning the initial condition.
On the other hand, if $f_a \ll 10^{12}$\,GeV, the axion abundance is insufficient unless the initial field value lies near the hilltop of the potential.
However, such fine-tuned initial conditions result in unacceptably large isocurvature perturbations~\cite{Lyth:1991ub,Kobayashi:2013nva}.

If the PQ symmetry is broken after inflation, on the other hand, the axion field value is randomly determined after the PQ phase transition, and cosmic strings are formed. This is called the post-inflationary scenario.
Then, later in the evolution of the universe, domain walls attached to the cosmic strings appear around the QCD crossover. If the string-wall network survives until the present time, its energy dominates the universe and spoils the success of the $\Lambda$CDM  cosmology. Thus, the string-wall network must decay in the early universe. 

The fate of the string-wall network depends on the domain wall number, $N_\mathrm{DW}$, which is the number of domain walls attached to one cosmic string. If $N_\mathrm{DW} = 1$, the network collapses soon after formation due to the tension of the domain walls. At the collapse, the network emits the QCD axions, which can account for dark matter~\cite{Davis:1986xc} {(see also Refs.\,\cite{Redi:2022llj,Gonzalez:2022mcx, Aldabergenov:2024fws} for a longer lifetime of domain walls due to inflationary dynamics}). If $N_\mathrm{DW} > 1$, the {tension of} domain walls attached to the strings balance with each other, and the network becomes stable. To avoid the overclosure of the universe, we need to introduce a potential bias, which makes the network unstable~\cite{Sikivie:1982qv}. However, such an additional potential generally shifts the potential minimum of the QCD axion and potentially spoils the PQ mechanism as a solution to the strong CP problem~\cite{Barr:1992qq,Kamionkowski:1992mf,Holman:1992us,Kawasaki:2014sqa,Chang:2023rll,Beyer:2022ywc}. 

As a broader class of particles, we can consider axion-like particles (ALPs), which are predicted in various extensions of the Standard Model, including string theory~\cite{Witten:1984dg,Conlon:2006tq,Svrcek:2006yi,Arvanitaki:2009fg,Cicoli:2012sz}.
ALPs share some properties with the QCD axion, such as a periodic potential and coupling to the Standard Model sector.
However, ALPs can have a wide range of masses and coupling strengths, in contrast to the QCD axion, whose scale is closely tied to the QCD scale.
If multiple axions are present, they can mix in the potential, and a specific linear combination of them could serve as the QCD axion. 
The mixing between the QCD axion and an ALP has been extensively studied in 
a various context such as the level crossing~\cite{Kitajima:2014xla,Daido:2015bva,Daido:2015cba,Ho:2018qur,Cyncynates:2023esj}, the misalignment mechanism~\cite{Murai:2023xjn}, decaying dark matter~\cite{Higaki:2014qua}, clockwork mechanism~\cite{Higaki:2015jag,Higaki:2016jjh,Higaki:2016yqk,Long:2018nsl}, and inflation~\cite{Takahashi:2019pqf,Kobayashi:2019eyg,Kobayashi:2020ryx,Narita:2023naj}.

In this paper, we discuss a novel phenomenon where the domain wall of a heavy ALP induces a domain wall-like structure in the QCD axion in a model where the two axions mix in the potential. 
For this realization, we impose different initial conditions for the QCD axion and the heavy ALP, which naturally depends on the UV completions of the axion and ALP, and the inflaton sector. For the QCD axion, we assume a pre-inflationary initial condition, resulting in a nearly spatially uniform initial condition. In contrast, for the heavy ALP, we consider a post-inflationary initial condition where spontaneous symmetry breaking leads to the formation of strings and domain walls. We also assume that the ALP is much heavier than the QCD axion. Due to the potential mixing,   the effective $\theta$ parameter takes different values in each domain separated by the ALP domain wall, shifting the potential minimum of the QCD axion in different domains. Consequently, despite imposing a pre-inflationary initial condition on the QCD axion, which typically leads to a spatially uniform configuration, a domain-wall-like structure of the QCD axion emerges after the QCD axion starts to oscillate. We refer to this structure as induced domain walls.
In contrast to {standard} 
domain walls, the tension of the induced domain wall depends not only on the potential of the QCD axion, but also on the position of the potential minimum in different domains. 

The ALP domain walls cause a cosmological domain wall problem~\cite{Zeldovich:1974uw}, so they need to decay. This can be achieved by introducing a potential bias in the heavy ALP potential. As a consequence, both domain wall networks eventually collapse while the QCD axion continues to solve the strong CP problem. The timing of the string wall network collapse depends on the magnitude of the potential bias.  To preserve the success of Big Bang Nucleosynthesis (BBN), the heavy ALPs must decay to Standard Model particles at temperatures $T \gtrsim 10\, \mathrm{MeV}$~\cite{Kawasaki:1999na,Kawasaki:2000en,Ichikawa:2005vw}. In addition, the collapse of the heavy ALP domain wall produces a large amount of gravitational waves, which can explain the background gravitational waves in the nHz frequency band suggested by recent observations of the pulsar timing array (PTA)~\cite{NANOGrav:2023gor,Antoniadis:2023ott,Reardon:2023gzh,Xu:2023wog}.\footnote{%
See, for instance, Refs.~\cite{Kitajima:2023cek,Bai:2023cqj,Blasi:2023sej,Gouttenoire:2023ftk,Ferreira:2024eru,Gruber:2024jtc} for the explanation of the PTA result using the collapse of axion domain walls.
In this context, the formation of primordial black holes with masses of order $M_{\odot}$ was first noted in Ref.~\cite{Kitajima:2023cek}. For a more detailed analysis, see Refs.~\cite{Gouttenoire:2023ftk,Ferreira:2024eru}. Note that the initial version of Ref.~\cite{Gouttenoire:2023ftk} suggested that PBH overproduction ruled out the domain wall interpretation of the PTA signal, but this conclusion was revised in the subsequent version.}
Moreover, as we will see later, the QCD axions produced by the collapse of the induced domain wall can explain all dark matter even with
$f_a = \mathcal{O}(10^{9})$\,GeV despite the pre-inflationary initial condition. 

Our mechanism is not restricted to the QCD axion. It is also possible to create induced domain walls for string axions or axions with a very large decay constant, which would normally be challenging since the corresponding symmetry is not considered to be restored\footnote{
It is possible in a certain setup where the initial position of the axion is set to be around the potential maximum~\cite{Takahashi:2019pqf}
} (see Appendix~\ref{app:2}).
We will also discuss the cosmological implications of such induced axion domain walls for cosmic birefringence~\cite{Takahashi:2020tqv,Kitajima:2022jzz,Gonzalez:2022mcx,Kitajima:2023kzu,Ferreira:2023jbu} ({see also the experimental papers {\cite{Minami:2020odp,Diego-Palazuelos:2022dsq,Eskilt:2022wav,Eskilt:2022cff,Cosmoglobe:2023pgf}}}).

The rest of this paper is organized as follows.
In Sec.~\ref{sec: model}, we describe our set-up.
Then, we explain the dynamics of the axions and, in particular, the formation of the induced domain walls in Sec.~\ref{sec: induced DW}.
We discuss the collapse of the domain walls and study the production of the QCD axion dark matter and gravitational waves in Sec.~\ref{sec: DW collapse}.
Sec.~\ref{sec: summary} is devoted to the summary and discussion of our results.

\section{Model}
\label{sec: model}
We consider two complex scalar fields, $A$ and $\Phi$, with corresponding two global symmetries, U(1)$_\mathrm{PQ}$ and U(1)$_\mathrm{H}$.\footnote{
Since we impose a pre-inflationary initial condition on the axion $a$, it is not necessary to introduce the complex scalar $A$ for our mechanism to work. In fact, we will apply our mechanism to string axions for which no such $A$ exists.
}
At low energies, both $A$ and $\Phi$ acquire nonzero vacuum expectation values (VEVs), and these symmetries are spontaneously broken.
Then, two axions, $a$ and $\phi$, emerge from the phases of $A$ and $\Phi$, respectively;
\begin{align}
    A &= \frac{f_a}{\sqrt{2}} e^{i \frac{a}{f_a}},\\
    \Phi  &= \frac{f_\phi}{\sqrt{2}} e^{i \frac{\phi}{f_\phi}},
\end{align}
where $f_a$ and $f_\phi$ are the axion decay constants, and we dropped the radial degrees of freedom. 
We assume that the axions acquire the potential through non-perturbative effects of QCD and the hidden sector.
In general, $a$ and $\phi$ can mix in the potential, which depends on the charge assignment of $A$ and $\Phi$.
Here, we consider the low-energy effective potential for $a$ and $\phi$ given by~\cite{Kitajima:2014xla,Daido:2015bva,Daido:2015cba,Ho:2018qur,Cyncynates:2023esj} 
\begin{align}
    V(a,\phi)
    &= V_\mathrm{mix}(a,\phi) 
    + V_\mathrm{DW}(\phi) 
    \nonumber \\
    &=
    \chi(T) \left[ 1 - \cos \left( N_a \frac{a}{f_a} + N_\phi \frac{\phi}{f_\phi} \right) \right]
    + \Lambda^4 \left[ 1 - \cos \left( N_\mathrm{DW} \frac{\phi}{f_\phi} \right) \right]
    \ ,
\end{align}
where $V_\mathrm{mix}$ comes from the non-perturbative effects of QCD, $V_\mathrm{DW}$ 
from that of the hidden sector, and $N_a$, $N_\phi$, and $N_{\rm DW}$ are integers. 
This potential can be obtained by coupling $A$ to Peccei-Quinn fermions that only generate an anomaly for QCD and $\Phi$ to fermions that generate anomalies for both QCD and the hidden gauge group. 
A very simple model giving the origin of the setup and solving the quality problem for the QCD axion is provided in Appendix~\ref{app:model1}.

Note that the definition of $f_a$ is different from the usual one by a factor of $N_a$. {For instance, the SN1987A bound on the QCD axion reads $f_a/N_a\gtrsim 10^8\GEV$ when $\phi$ is much heavier than $a$.}
We also set the origin of $a$ and $\phi$ so that {the potential is minimized at}
$a = \phi = 0$.

The topological susceptibility of QCD, $\chi(T)$, depends on the temperature as 
\begin{align}
    \chi(T)
    \equiv 
    \frac{m_a^2(T) f_a^2}{N_a^2}
    =
    \left\{
        \begin{array}{ll}
            \chi_0 & \quad (T < T_\mathrm{QCD})
            \\
            \chi_0 \left( \frac{T}{T_\mathrm{QCD}} \right)^{-n} & \quad (T \geq T_\mathrm{QCD})
        \end{array}
    \right.
    \ ,
\end{align} 
where we adopt $\chi_0 \simeq (75.6\,\mathrm{MeV})^4$, $T_\mathrm{QCD} \simeq 153\,\mathrm{MeV}$, and $n \simeq 8.16$~\cite{Borsanyi:2016ksw}. Note that $m_a(T)$ represents the mass of $a$ when we fix $\phi = 0$ , or equivalently when we integrate out $\phi$ assuming it is much heavier than $a$
For later convenience, we also define {$m_{a0} \equiv m_a(T < T_\mathrm{QCD})$ and} 
$m_\phi \equiv N_{\rm DW}\Lambda^2/f_\phi$
, where $m_\phi$ is the mass of $\phi$ arising from $V_\mathrm{DW}$ around its minima, and it is approximately equal to the mass of $\phi$ at temperatures $T \gg T_{\rm QCD}$.
Throughout this paper we assume $\phi$ is much heavier than $a$, $m_\phi \gg m_{a0}$.

In this model, 
only $V_\mathrm{mix}$ depends on $a$ in the {two} terms of $V(a,\phi)$. Since we assume that $\phi$ is much heavier than $a$,
it is $a$ that becomes the QCD axion in the low energy, solving the strong CP problem.

\section{Induced domain walls}
\label{sec: induced DW}

If a complex scalar field with a global U(1) symmetry acquires a nonzero VEV before or during inflation, the corresponding axion becomes nearly spatially homogeneous after inflation.
On the other hand, if spontaneous symmetry breaking (SSB) occurs after inflation, the axion field takes spatially random values, and cosmic strings form in association with the SSB of the U(1) symmetry.
In this scenario, when the axion potential develops and drives the axion towards its minima, domain walls emerge, attached to the cosmic strings. We assume these different initial conditions for $A$ and $\Phi$, respectively. 
Models that naturally realize these different initial conditions are presented in Appendix \ref{app:2}.
The comprehensive study with more generic choices and initial conditions will be given elsewhere (see e.g. Ref.\,\cite{Lee:2024toz}).

In the following, we assume that $A$ acquires a nonzero VEV before or during inflation, consistent with the pre-inflationary scenario. On the other hand, $\Phi$ is assumed to acquire a nonzero VEV after inflation, aligning with the post-inflationary scenario.
Consequently, $a$ becomes nearly homogeneous after inflation, while $\phi$ forms cosmic string configurations following SSB.
When $m_\phi$ exceeds the Hubble parameter, domain walls of $\phi$ form, and $N_{\rm DW}$ walls are attached to each string. Since we are interested in long-lived domain walls, we assume $N_\mathrm{DW} \geq 2$ in the following analysis.
We further assume a hierarchy in the mass and the tension,
\begin{align}
\label{eq:hierarchy1}
    m_\phi & \gg m_{a0},\\
\label{eq:hierarchy2}
    \frac{m_\phi f_\phi^2}{N_\mathrm{DW}^2} &\gg \frac{m_{a0} f_a^2}{N_a^2},
\end{align} 
for simplicity (See discussions below \Eq{astring} for the other cases).
Under these conditions, $\phi$ domain walls form first while $a$ remains constant. We can also neglect the backreaction from $a$ to $\phi$, 
namely, the $\phi$ domain wall dynamics and configuration are not affected by $V_{\rm mix}$.

As the cosmic temperature drops down to the QCD scale, $V_{\rm mix}$ grows, and the QCD axion $a$ begins to oscillate around the potential minimum. However, in the presence of $\phi$ domain walls, the potential minimum of $a$ shifts across the $\phi$ wall. This induces a domain wall-like structure for $a$, even though we assume the pre-inflationary initial condition for $a$.

To see the solution of $a$ in the presence of $\phi$ domain walls, we consider a simplified situation in which
{a $\phi$ domain wall coincides with the $xy$ plane, neglecting the cosmic expansion.}
Since we have assumed $m_\phi \gg m_{a0}$, the domain wall width ($\sim m_\phi^{-1}$) is much smaller than the scale of the spatial dependence of $a$.
Here, we approximate $\phi = 0$ for $z < 0$ and $\phi = \Delta \phi \equiv 2\pi f_\phi/N_\mathrm{DW}$ for $z > 0$.
Then, the potential for $a$ is effectively given by
\begin{equation}
    V_a(a) 
    =
    \left\{
    \begin{array}{cc}
        \displaystyle{\frac{m_a^2 f_a^2}{N_a^2} }
        \left[ 1 - \cos \left( N_a \frac{a}{f_a} \right)\right]
        & (z < 0)
        \\
        &\\
        \displaystyle{\frac{m_a^2 f_a^2}{N_a^2} }
        \left[ 1 - \cos \left( N_a \frac{a}{f_a} + \frac{2\pi N_\phi}{N_\mathrm{DW}} \right)\right]
        & (z > 0)
    \end{array}
    \right.
    \ ,
\end{equation}
and the equation of motion for $a$ becomes
\begin{align}
    \partial_t^2 a(t,z) - \partial_z^2 a(t,z) 
    + \frac{m_a^2 f_a}{N_a} \sin \left( N_a\frac{a(t,z)}{f_a} \right)
    &= 0
    \quad (z < 0) 
    \ ,
    \\
    \partial_t^2 a(t,z) - \partial_z^2 a(t,z)
    + \frac{m_a^2 f_a}{N_a} \sin 
    \left(
        N_a\frac{a(t,z)}{f_a} + \frac{2\pi N_\phi}{N_\mathrm{DW}} 
    \right)
    &= 0
    \quad (z > 0)
    \ .
\end{align}
Since we are interested in a time scale much shorter than the cosmic expansion,  we neglect the Hubble friction term and the temperature dependence of $m_a$.

First, we focus on each domain far from the domain wall.
By shifting $a \to a + 2\pi f_a j/N_a$ with $j$ being an integer, we can always redefine $a$ so that the initial value of $a$, $a_\mathrm{ini}$, satisfies $-\pi f_a/N_a < a_\mathrm{ini} \leq \pi f_a/N_a$.
Then, for $z < 0$, $a$ starts to oscillate around $a = 0$ with an amplitude of $|a_\mathrm{ini}|$.
For $z \geq 0$, $a$ starts to oscillate around $a = a_\mathrm{min}$ with an amplitude of $|a_\mathrm{ini} - a_\mathrm{min}|$, where $a_\mathrm{min}$ is the minimum of {$V_a$} {(for $z \geq 0$)} nearest to $a_\mathrm{ini}$, {and it is} given by
\begin{align}
    a_\mathrm{min}
    =
    2 \pi f_a \left( -\frac{N_\phi}{N_a N_\mathrm{DW}} + \frac{l}{N_a} \right)
    \ ,
\end{align}
where $l$ is an integer.
Note that $-2 \pi < N_a a_\mathrm{min}/f_a < 2 \pi$.
We show a schematic representation of $V_a$ for $z>0$ (red solid) and $z<0$ (blue dashed) in Fig.~\ref{fig: Va}.
\begin{figure}[!t]
    \begin{center}  
        \includegraphics[width=0.8\textwidth]{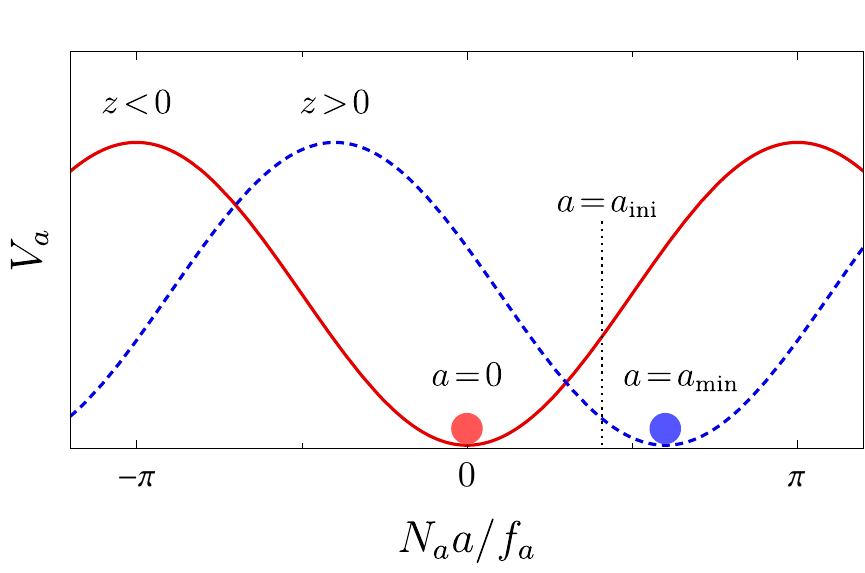}
        \end{center}
    \caption{%
        The red-solid and blue-dashed lines show $V_a(a)$ for $z<0$ and $z \geq 0$, respectively.
        Depending on the initial condition, $a = a_\mathrm{ini}$, $a$ begins to oscillate around different minima: $a = 0$ for $z<0$ and $a = a_\mathrm{min}$ for $z \geq 0$.
        The vertical dotted line shows an example value of $a_\mathrm{ini}$, and the red and blue circles show the corresponding minima.
    }
    \label{fig: Va} 
\end{figure}
Since $a$ oscillates around different minima in different domains, the field value of $a$ transits across the domain wall. As a result, the gradient and potential energies are localized around the domain wall of $\phi$.
In this sense, the domain wall of $\phi$ induces a domain wall of $a$ via $V_\mathrm{mix}$, despite we have assumed a homogeneous initial condition for $a$.
{Note that, if $N_\phi/N_\mathrm{DW}$ is an integer, {we have $a_\mathrm{min} = 0$, and thus, the two minima are aligned along the $\phi$ direction.} In this case,
the induced domain walls are not formed when neglecting the width of the $\phi$ walls.}\footnote{{If $m_\phi$ is small, which is not the main focus of this paper, the domain wall solution evolves to span not only the $\phi$ direction but also the $a$ direction, although it might be difficult to separate the induced walls from the original $\phi$ domain walls. In this case, some amount of the QCD axion could be generated from the domain wall collapse.}}

Next, we study the configuration of the induced domain wall.
A {static} solution, $\bar{a}(z)$, follows the equations of motion:
\begin{align}
    \partial_z^2 \bar{a}(z) 
    &\simeq
    \frac{m_a^2 f_a}{N_a} 
    \sin \left(N_a \frac{\bar{a}(z)}{f_a} \right)
    &(z < 0) 
    \ ,
    \\
    \partial_z^2 \bar{a}(z)
    &\simeq
    \frac{m_a^2 f_a}{N_a} 
    \sin \left(N_a \frac{\bar{a}(z) - a_\mathrm{min}}{f_a} \right)
    &(z > 0)
    \ ,
\end{align}
with the boundary conditions of 
\begin{align}
    \bar{a}(z \to -\infty)
    & = 0
    \ ,
    \\
    \bar{a}(z \to \infty)
    & = a_\mathrm{min}
    \ .
\end{align}
We show the $z$-dependence of the axion fields and the energy density of $a$ for $|N_a a_\mathrm{min}/f_a| \ll 1$ in  Fig.~\ref{fig: DW configuration}.
\begin{figure}[!t]
    \begin{center}  
        \includegraphics[width=0.8\textwidth]{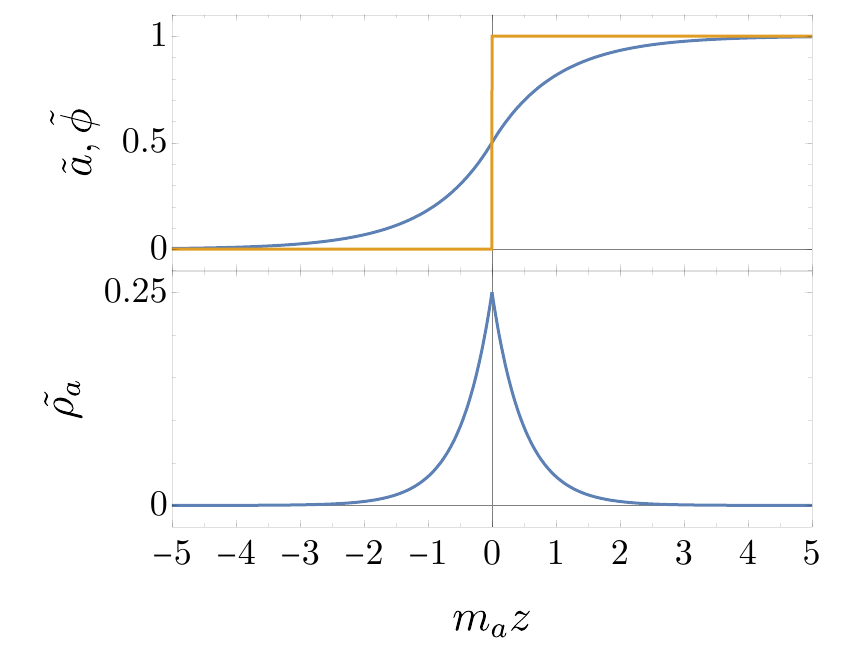}
        \end{center}
    \caption{%
        Dependence of the axion fields and the energy density of $a$ on $z$ for $|N_a a_\mathrm{min}/f_a| \ll 1$.
        We show the dimensionless quantities defined by $\tilde{a} \equiv \bar{a}/a_\mathrm{min}$ (blue) and $\tilde{\phi} \equiv \phi/\Delta \phi$ (orange) in the upper panel, and $\tilde{\rho}_a \equiv \rho_a/(m_a^2 a_\mathrm{min}^2)$ in the lower panel.
    }
    \label{fig: DW configuration} 
\end{figure}
Here, the energy density of $a$ is given by 
\begin{align}
    \rho_a(z) 
    &\equiv
    \frac{1}{2} \left( \partial_z \bar{a} \right)^2
    +
    V_a(a)
    \ .
\end{align}
{
One can see that the induced domain wall appears around the $\phi$ wall as expected, and that the energy density of $a$ is localized around the $\phi$ wall. 
}

The tension of the induced domain wall is evaluated as
\begin{align}
    \sigma_a 
    &\equiv
    \int_{-\infty}^{\infty} \mathrm{d} z \, 
    \rho_a(z)
    \ .
\end{align}
Although the shape of the domain wall depends on $a_\mathrm{min}$, we expect $\sigma \sim m_a a_\mathrm{min}^2$ from the maximum potential value $\sim m_a^2 a_\mathrm{min}^2$ and the wall width $\sim m_a^{-1}$.
Thus, we introduce a dimensionless parameter $\kappa$ defined by 
\begin{align}
    {\sigma_a
    =
    \kappa m_a a_\mathrm{min}^2}
    \ .
\end{align}
We show the dependence of the tension on $N_a a_\mathrm{min}/f_a$ in Figs.~\ref{fig: tension} and \ref{fig: tension kappa}.
In Fig.~\ref{fig: tension} we show the domain wall tension normalized by $\sigma_{a,\mathrm{max}} \equiv 8 m_a f_a^2/N_a^2$, which corresponds to the domain wall tension for the cosine potential.
\begin{figure}[!t]
    \begin{center}  
        \includegraphics[width=0.8\textwidth]{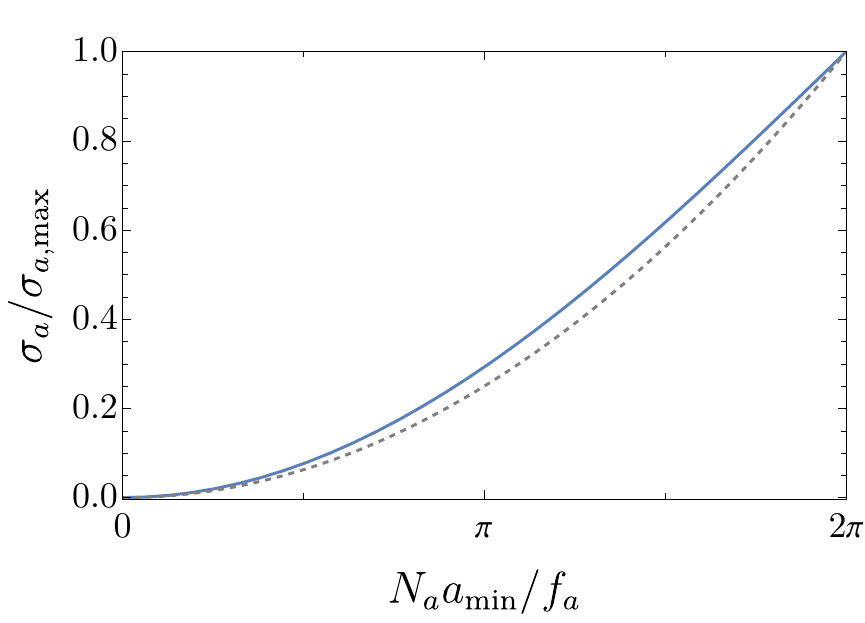}
        \end{center}
    \caption{%
        Tension of the induced domain wall normalized by the maximum value of the tension, $\sigma_{a,\mathrm{max}} \equiv 8 m_a f_a^2/N_a^2$, which corresponds to the domain wall tension in the conventional case.
        The gray-dashed line is a reference line $\propto (N_a a_\mathrm{min}/f_a)^2$.
    }
    \label{fig: tension} 
\end{figure}
\begin{figure}[!t]
    \begin{center}  
        \includegraphics[width=0.8\textwidth]{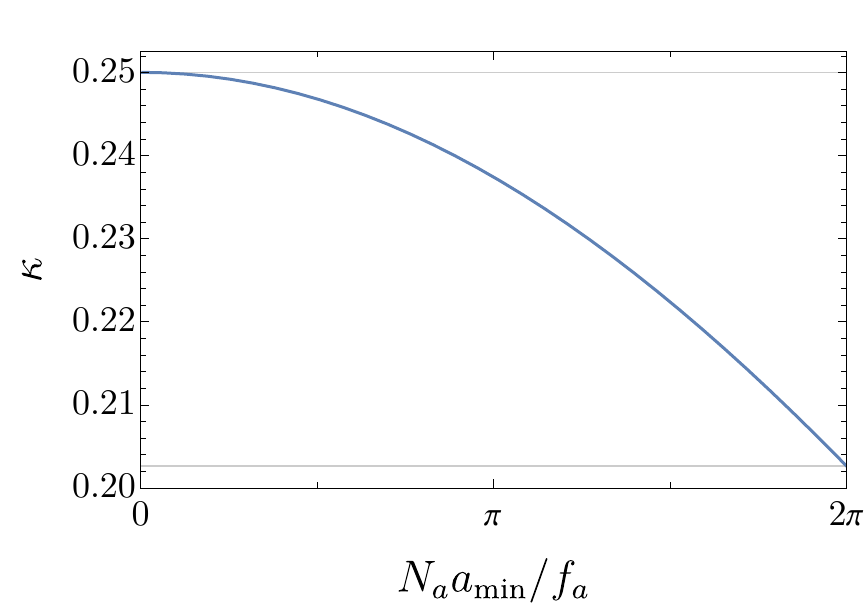}
   \end{center}
    \caption{%
     The parameter $\kappa$ that parametrizes
        the tension of the induced domain wall.
        The horizontal gray lines show $\kappa = 1/4$ and $2/\pi^2$, which correspond to the 
        {limit of the quadratic and cosine potentials,}
        respectively.}
    \label{fig: tension kappa} 
\end{figure}
{We see that the tension roughly follows $\sigma_a \propto a_\mathrm{min}^2$.
Thus, in our scenario, the tension of the induced domain wall is generically smaller than in the conventional case, }$\sigma_{a,\mathrm{max}} = 8 m_a f_a^2/N_a^2$.
In Fig.~\ref{fig: tension kappa}, we show the dependence of $\kappa$ on the $N_a a_\mathrm{min}/f_a$.
Note that {$\kappa$} approaches ${\kappa} \simeq 1/4$ in the limit of $|N_a a_\mathrm{min}/f_a| \ll 1$, while it tends to
${\kappa} \simeq 2/\pi^2$ when $|N_a a_\mathrm{min}/f_a| \simeq 2\pi$.
See Appendix~\ref{app: DW tension} for more details on the configuration and tension of the induced domain wall.

We have discussed the static solution, $\bar{a}(z)$, and found that the domain wall of $a$ is induced along that of $\phi$.
In a realistic situation, the $\phi$ domain walls follow the scaling solution, and they move around and annihilate each other when they collide. As the $\phi$ domain walls move around, the induced domain walls are considered to follow the motion.
When the induced domain wall passes a certain point in space, the potential minimum of $a$ suddenly changes there. It is not obvious how the motion of the induced domain wall affects the dynamics of $a$. We then consider the time evolution of $a(t,z)$ around the domain wall and show that the motion of the induced domain walls does not significantly affect the axion dynamics.

To this end, we decompose the axion field $a$ into the static solution and {a perturbation around it} as
\begin{align}
    a(t,z)
    =
    \bar{a}(z) + \delta a (t, z)
    \ .
\end{align}
Then, the equation of motion for $\delta a$ is given by 
\begin{align}
    \left[
        \partial_t^2 - \partial_z^2
    + m_a^2 \cos \left( N_a\frac{\bar{a}}{f_a} \right) 
    \right] 
    \delta a
    &= 0
    \quad (z < 0) 
    \ ,
    \\
    \left[
        \partial_t^2 - \partial_z^2
    + m_a^2 \cos \left( N_a\frac{\bar{a}- a_\mathrm{min}}{f_a} \right) 
    \right] 
    \delta a
    &= 0
    \quad (z > 0)
    \ ,
\end{align}
where we assumed $|N_a \delta a/f_a| \ll 1$.
If $|N_a a_\mathrm{min}/f_a| \ll 1$, the equation of motion is approximately given by
\begin{align}
    \left[ 
        \partial_t^2 - \partial_z^2 + m_a^2 
    \right] 
    \delta a(t,z)
    &\simeq 0
    \ ,
\end{align}
for all $z$.
Thus, the perturbation behaves as a superposition of freely propagating plane waves, which implies that axion perturbations are neither enhanced nor suppressed when passing through the domain walls.
In particular,  an incoming wave propagating toward the positive $z$ direction simply passes through the domain wall, without reflection.
If $|N_a a_\mathrm{min}/f_a| = \mathcal{O}(1)$, the mass squared of $\delta a$, $m_{a,\mathrm{eff}}^2(z)$, deviates from $m_a^2$ around the wall. 
{Specifically, if $|N_a a_\mathrm{min}/f_a| > \pi$, the mass squared becomes negative around the domain wall.}
This phenomenon, however, merely increases the wavenumber and does not affect the transmission rate, as the effective potential for $\delta a$ remains continuous everywhere. 
Thus, we 
conclude that the existence and motion of the induced domain wall do not 
affect the time evolution of the axion excitation.\footnote{This scenario contrasts with the so-called bubble misalignment mechanism~\cite{Lee:2024oaz}, where the axion mass varies across the bubble wall while the location of the potential minimum remains the same.} 
We also note that this discussion does not hold with $|\d a N_a/ f_a|=\O(1)$.

We have thus far examined individual induced domain walls. Given our assumption of a post-inflationary initial condition for $\phi$, $N_{\rm DW}$ domain walls of $\phi$ are attached to {each} cosmic string. Fig.~\ref{fig:string-wall} illustrates the field configuration around a cosmic string,
specifically when tracing a path that encircles the string at a sufficient distance,
for the case where $N_\mathrm{DW} = 2$ and $N_a = N_\phi = 1$.
In this figure, we identify $a/f_a \sim a/f_a + 2\pi$ and $\phi/f_\phi \sim \phi/f_\phi + 2\pi$. The horizontal {dashed} lines (red) denote the potential minima of $V_{\rm DW}(\phi)$, while the diagonal dashed line (black) denotes the potential minimum of $V_{\rm mix}(a,\phi)$.
The string configuration before $a$ begins to oscillate is represented by a vertical thin solid line ({green}). After the onset of $a$ oscillations, the configuration is represented by a thick solid line (blue) composed of vertical and horizontal segments. Here,
due to the assumed mass hierarchy, the $\phi$ walls are represented by vertical line segments, while the induced walls are shown as horizontal segments. The two vacua are indicated by {purple} stars. 
One can see that these vacua are connected by walls, with $\phi$ walls sandwiched between the induced walls. See also Fig.~\ref{fig: DW configuration}.

For $N_{\rm DW} \geq 3$, the field configuration around the string becomes more complex but maintains a similar structure. The potential minima of $a$ can be written as
\begin{align}
   a_{{\rm{min}},k}
    =
    2 \pi f_a \left( -\frac{k N_\phi}{N_a N_{\rm DW}} + \frac{l_k}{N_a} \right)
\end{align}
 where $k = 0, \ldots, N_\mathrm{DW} - 1$ labels the minima of $V_\phi$ as $\phi_k = 2\pi k f_\phi/N_{\rm DW}$, 
 and $l_k$ is an integer that labels the minimum of $V_{\rm mix}$ closest to $(a,\phi) = (a_{\rm ini},\phi_k)$.  
 In particular, $a_\mathrm{min,0} = 0$ and $a_\mathrm{min,1} = a_\mathrm{min}$. The adjacent minima, $a_{{\rm{min}},k}$ and $a_{{\rm{min}},k+1}$, are connected by the $\phi$ wall and induced walls as in Fig.~\ref{fig:string-wall}. Note that the induced walls go back and forth along the $a$ direction since there is no winding number along the $a$ direction.

When $N_{\rm DW}>2$, we generically have more than two induced walls around a single $\phi$ string.
In this case, one of the induced walls has a tension larger than the others.
For example, let us consider the parameter set of $N_\phi=N_a=1$, $N_{\rm DW}=3$, from which we get $a_{{\rm min},1}=-2\pi f_a/3$, $a_{{\rm min},2}=2\pi f_a/3$.
The tension of the domain wall connecting $a_{{\rm min},1}$ and $a_{{\rm min},2}$ is estimated to be approximately four times larger than that of the other walls, as the tension scales with the square of the distance between minima (see Fig.~\ref{fig: tension}).

\begin{figure}[!t]
    \begin{center}  
        \includegraphics[width=0.6\textwidth]{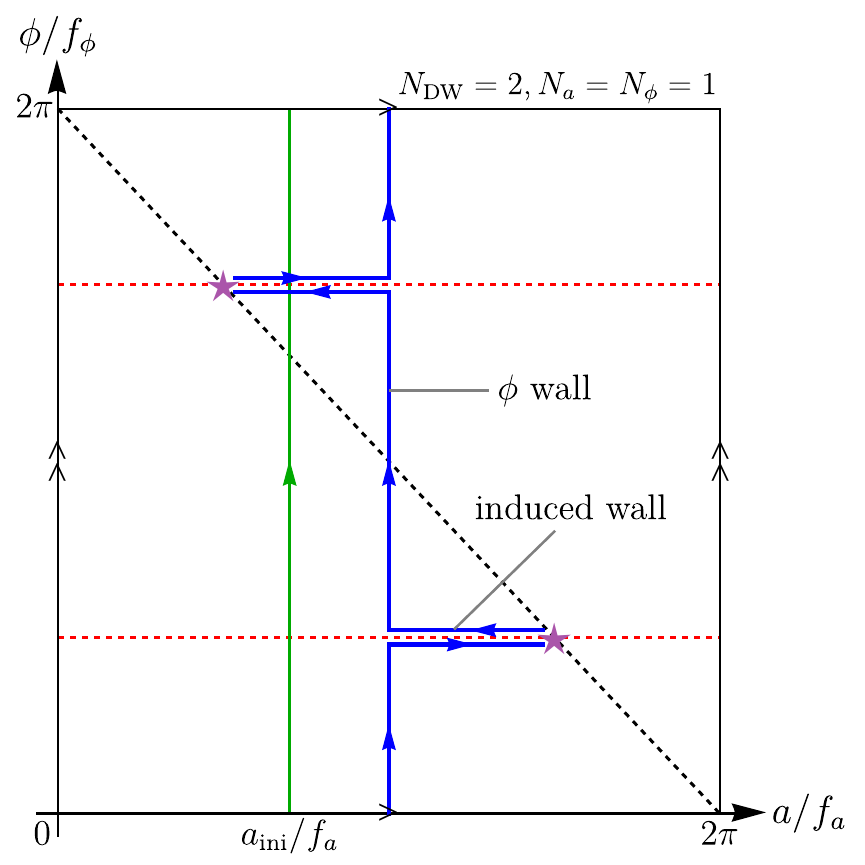}
    \end{center}
    \caption{%
  The field configuration of the axions around the cosmic string of $\phi$  in the case of $N_\mathrm{DW} = 2$ and $N_a = N_\phi = 1$. The winding numbers of $a$ and $\phi$ are $0$ and $1$, respectively, reflecting their initial condition. {The red and black dashed lines represent the minima of $V_{\rm DW}(\phi)$ and $V_{\rm mix}(a,\phi)$, respectively.} The vertical line segments correspond to the heavy $\phi$ walls, while the horizontal ones represent the induced domain walls of $a$. {The green vertical line at $a=a_\mathrm{ini}$ is an example of the initial field configuration before the QCD {crossover}, and the blue line segments show one stable solution after {the crossover}.}
   {For illustrative purposes, we shift the potential minima of $V_\mathrm{DW}$ by $\pi f_\phi/4$.} 
  }
    \label{fig:string-wall} 
\end{figure}

The domain wall configuration shown in Fig.~\ref{fig:string-wall} represents the field structure at a large distance from a string. Within a radius of order $1/m_{a0}$ from the string core, the gradient energy of $a$ becomes dominant, preventing the axion field $a$ from stabilizing at the minimum in each domain. Consequently, we expect the axion $a$ to take on a value close to the average of the potential minima:\footnote{Note also that the mixing term $V_{\rm mix}$ is expected to be suppressed in the vicinity of the $\phi$ string core.}
\begin{align}
\left.a \right|_{\rm near~strings} \sim \langle a \rangle
= \frac{1}{N_{\rm DW}} \sum_k a_{{\rm min},k}.\laq{astring}
\end{align}
This implies that the induced domain walls are not directly attached to the $\phi$ strings, which is consistent with the absence of a winding number along the $a$ direction.

Thus far, we have assumed a hierarchy both in the axion masses and domain wall tensions, Eqs.\,{(\ref{eq:hierarchy1}) and (\ref{eq:hierarchy2}), for simplicity. In fact, these assumptions are somewhat stronger than what is really needed for our consideration, and they can be relaxed to some extent.}
Since we are interested in the formation of domain walls induced by {pre-existing} domain walls, we require that $\phi$-domain walls are already formed before $a$ begins to oscillate.
This {is the case} if $m_\phi > m_a(T_{\rm osc})$, where $T_{\rm osc}$ is the temperature of the universe {when $a$ begins to oscillate, i.e., } $H(T_{\rm osc}) = m_a(T_{\rm osc})$. This condition, $m_\phi > m_a(T_{\rm osc})$, is weaker than the condition  (\ref{eq:hierarchy1}). 
For instance, $m_{a0}$ can be larger than $m_\phi$ while this condition is satisfied, in which case the width of the induced domain walls will likely be determined by the width of the $\phi$ domain wall.
On the other hand, the hierarchy in the tension, Eq.\,(\ref{eq:hierarchy2}), ensures that $V_{\rm mix}$ does not act as a potential bias for the $\phi$ domain walls until $H \sim m_{a0}$, when the spatial configuration of the induced domain walls reaches its final form. 
It would be interesting and worthy of further study to classify all the possible dynamics and the field configuration when this assumed hierarchy in the tension is relaxed. We leave it for future work.

Instead, let us briefly comment on the case where $m_\phi f_\phi^2/N_{\rm DW}^2$ and $m_a(T_{\rm osc}) f_a^2/N_a^2$ are comparable or even $m_\phi f_\phi^2/N_{\rm DW}^2 < m_a(T_{\rm osc}) f_a^2/N_a^2$. We continue to assume that $N_\phi/N_{\rm DW}$ is not an integer for the formation of the induced domain walls. Before the onset of $a$ oscillations, the potential $V_{\rm mix}$ behaves as a potential bias for the $\phi$ walls. Let us define the decay temperature $T_{\rm dec'}$ by $m_\phi f_\phi^2 H(T_{\rm dec'})/N_{\rm DW}^2 = m_a^2(T_{\rm dec'}) f_a^2/N_a^2$. Then, if $T_{\rm dec'} \gg T_{\rm osc}$, the $\phi$ string-wall network collapses before inducing the $a$-domain walls. This is nothing but the scenario studied in Ref.\,\cite{Kitajima:2023cek}, where the onset of the oscillations of the QCD axion $a$ is delayed due to
a very large decay constant $f_a\gg 10^{12}\,\GEV$. In this case, the induced domain walls are not formed since the $\phi$ domain walls collapse early.
Thus,  the QCD axion $a$ is not produced from the domain wall dynamics.

On the other hand,  if $T_{\rm dec'} \lesssim T_{\rm osc}$, the potential bias gets suppressed since the axion $a$ starts to oscillate, and $T_{\rm dec'}$ no longer represents the decay temperature. However, if the two temperatures are not very different, the string-wall network is slightly biased toward one of the domains due to the suppressed potential bias, which results in the population bias after inducing the $a$ domain walls.   The population bias causes the $\phi$ string-wall network (as well as induced domain walls) to collapse. It is known that even a tiny population bias will cause domain walls to decay  quickly~\cite{Lalak:1993bp,Lalak:1994qt,Coulson:1995uq,Coulson:1995nv,Larsson:1996sp,Correia:2014kqa,Correia:2018tty}\footnote{
The robust stability of domain walls against the population bias in the case of inflationary initial fluctuations was overlooked in the literature. See Refs.~\cite{Gonzalez:2022mcx, Kitajima:2023kzu}.
} unless the domain wall network has correlations at superhorizon  scales~\cite{Gonzalez:2022mcx, Kitajima:2023kzu}. Thus, in this case, 
the domain wall problem could be evaded without introducing an extra potential bias for the $\phi$ domain walls.

Lastly, we evaluate the axion abundance contributed by the coherent oscillations.
Considering that each domain has a similar volume fraction, we can evaluate the contribution of the coherent oscillations to the axion abundance as
\begin{align}
    \frac{\rho_a(T)}{s(T)}
    &=
    m_{a0} \frac{n_a(T)}{s(T)}
    \nonumber \\
    &=
    m_{a0} \frac{n_{a,\mathrm{osc}}}{s(T_\mathrm{osc})}
    \nonumber \\
    &\simeq 
    \frac{m_{a0} m_{a,\mathrm{osc}}}{N_\mathrm{DW} s(T_\mathrm{osc})}
    \sum_{k=0}^{N_\mathrm{DW}-1} (a_\mathrm{ini}-a_{\mathrm{min},k})^2
    \ ,
\end{align}
where $s$ is the total entropy density, $\rho_a$ and $n_a$ are the energy and number densities of $a$, respectively, and the subscript ``osc'' denotes the quantities when $a$ starts to oscillate.
If the axion lies near the hilltop in some domain, the onset of oscillations is delayed there, and we need to include the anharmonic effect, which is neglected here for simplicity.
In the minimal setup with $N_\mathrm{DW} = 2$ and $N_a = N_\phi = 
1$, 
the axion abundance becomes
\begin{align}
    \frac{\rho_a(T)}{s(T)}
    &\simeq 
    \frac{m_{a0} m_{a,\mathrm{osc}}}{s(T_\mathrm{osc})}
    \frac{a_\mathrm{ini}^2 + (\pi f_a - |a_\mathrm{ini}|)^2}{2}
    \nonumber \\
    &\equiv 
    \frac{m_{a0} m_{a,\mathrm{osc}}}{s(T_\mathrm{osc})}
    \mathcal{F}(a_\mathrm{ini})
    \ .
\end{align}
Here, we parameterized the dependence on $a_\mathrm{ini}$ by $\mathcal{F}(a_\mathrm{ini})$, which satisfies
\begin{align}
    \frac{\pi^2 f_a^2}{4}
    \leq 
    \mathcal{F}(a_\mathrm{ini})
    \leq
    \frac{\pi^2 f_a^2}{2}
    \ .
    \label{eq: coherent min & max}
\end{align}
The axion abundance resulting from this coherent oscillation should be added to the axion abundance resulting from the collapse of the induced domain wall, which will be evaluated in the next section.

\section{Domain wall collapse}
\label{sec: DW collapse}

We have discussed the formation of induced domain walls and the axion abundance contributed by the coherent oscillations.
To avoid overclosure of the universe, the domain walls must collapse, which also contributes to the axion abundance.

In our scenario, one cosmic string of $\Phi$ has $N_\mathrm{DW}$ domain walls attached to it, and each domain wall is a superposition of domain walls of $a$ and $\phi$.
We label the domain wall separating $a = a_{\mathrm{min},k-1}$ and $a_{\mathrm{min},k}$ as the $k$-th wall, where $k = 1, \ldots , N_\mathrm{DW}$ and $a_{\mathrm{min},N_\mathrm{DW}} \equiv a_{\mathrm{min},0}$.
Since we consider $N_\mathrm{DW} \geq 2$, this string-wall system is long-lived.
Then, the number density of the strings and walls follows the scaling law. 
In other words, each Hubble volume has $\mathcal{O}(1)$ strings and walls.
In the scaling regime, the energy densities of the strings and walls evolve as
\begin{align}
    \rho_\mathrm{str} 
    \sim 
    \mu H^2
    \ , \quad 
    \rho_\mathrm{wall} 
    \sim
    \sigma H
    \ ,
\end{align}
where $\mu$ and $\sigma$ are the tensions of the strings and walls, and $H$ is the Hubble parameter.
Thus, $\rho_\mathrm{wall}$ comes to dominate the energy density of the string-wall network at some time.
After that, we can consider the dynamics of the network neglecting the strings.

To destabilize the domain walls, one possibility is to consider the dynamically generated population bias, $V_{\rm mix}$, before the onset of oscillations of the QCD axion, as mentioned previously.\footnote{This does not work when $N_\phi/N_{\rm DW}$ is an integer because the $V_{\rm mix}$ does not generate a potential bias for $\phi$. In this case, however, the $a$ domain walls are not induced in the thin wall limit of $\phi$ (see Fig.~\ref{fig: tension}). Thus, the QCD axions are not effectively produced by the domain wall collapse.}
 However, we expect that this is not effective, since the population bias is still very small at the onset of oscillations of the QCD axion for $f_a/N_a \lesssim 10^{12} \text{GeV}$. Therefore, we introduce a potential bias along the $\phi$ direction given by
\begin{align}
    V_\mathrm{bias}(\phi)
    &\equiv
    \epsilon \Lambda^4 \left[ 1 - \cos \left( N_\mathrm{b} \frac{\phi}{f_\phi} + \theta \right) \right]
    \ .
\end{align}
Here, we assume $\epsilon \ll 1$ {so that} the bias term does not affect the formation of domain walls discussed above.
Note that this bias term does not spoil the PQ solution to the strong CP problem, since $a$ remains light.
{Then}, the $k$-th domain wall feels pressure determined by the difference in the potential energy between both sides of the wall as
\begin{align}
    \Delta V_k
    &=
    |V_\mathrm{bias}(\phi_{k-1})
    -
    V_\mathrm{bias}(\phi_k)|
    \nonumber \\
    &=
    \epsilon \Lambda^4
    \left|
        \cos\left(N_\mathrm{b}\frac{\phi_{k-1}}{f_\phi} + \theta \right)
        -
        \cos\left(N_\mathrm{b}\frac{\phi_{k}}{f_\phi} + \theta \right)
    \right|
    \ ,
\end{align}
where $\phi_k \equiv 2\pi k f_\phi/N_\mathrm{DW}$.
When this pressure overcomes the domain wall tension {force}, the network collapses emitting axions.
We denote the typical bias by $\langle \Delta V \rangle$, while the bias depends on the parameters including the domain wall numbers in a complicated way.

{The dynamics of the string-wall network is the same as in the case of a string-wall network with a single ALP, since we assume the hierarchy in the tension (\ref{eq:hierarchy2}).}
The total tension of the superposed domain walls is given by
\begin{align}
    \sigma_\mathrm{tot}
    \simeq 
    \sigma_\phi
    \equiv 
    \frac{8 m_\phi f_\phi^2}{N_\mathrm{DW}^2}
    \ .
\end{align}
Consequently, the condition for the network collapse is 
\begin{align}
    \rho_{\phi,\mathrm{DW}}(t_\mathrm{dec})
    =
    \frac{\mathcal{A} \sigma_\phi}{t_\mathrm{dec}}
    \simeq 
    \langle \Delta V \rangle
    \ ,
\end{align}
where $\mathcal{A} = {\cal O}(1)$ is an area parameter describing the number of domain walls in each Hubble volume, and $t_\mathrm{dec}$ is the decay time of the network.
Although $\mathcal{A}$ in general {evolves in time and depends on the initial fluctuations of $\phi$}~\cite{Kitajima:2023kzu}, {in the following,} we set $\mathcal{A} = 1$ for simplicity.

In the following, we adopt $N_\mathrm{DW} = 2$ and $N_a = N_\phi = N_\mathrm{b} = 1$ as the minimal case.
Then, we obtain
\begin{align}
    \langle \Delta V \rangle 
    &=
    \Delta V_1
    =
    \Delta V_2
    \nonumber \\
    &=
    2 \epsilon \Lambda^4 |\cos\theta|
    \equiv
    c_\mathrm{b} \epsilon \Lambda^4
    \ ,
\end{align}
where $c_\mathrm{b} = 2 |\cos \theta|$ is an $\mathcal{O}(1)$ parameter of the potential bias.
Thus, we can evaluate $t_\mathrm{dec}$ as
\begin{align}
    t_\mathrm{dec}
    =
    \frac{1}{2H_\mathrm{dec}}
    \simeq 
    \frac{\sigma_\phi}{c_\mathrm{b} \epsilon \Lambda^4}
    =
    \frac{8}{c_\mathrm{b} \epsilon m_\phi}
    \ .
\end{align}
{We note that the decay time estimated in this way is shorter than that obtained from numerical simulations by a factor of $\O(1)$~\cite{Kawasaki:2014sqa, Kitajima:2023cek}, as it takes a finite time for the domain walls to decay completely.}
From the relation during the radiation-dominated era, $H = 1/(2t)$, and the Friedmann equation, we obtain
\begin{align}
    H_\mathrm{dec}
    &=
    \frac{c_\mathrm{b} \epsilon m_\phi}{16}
    \ , 
    \\
    T_\mathrm{dec}
    &=
    \left( \frac{90}{\pi^2 g_*(T_\mathrm{dec})} \right)^{1/4}
    \sqrt{\frac{c_\mathrm{b} \epsilon M_\mathrm{Pl} m_\phi}{16}}
    \nonumber \\
    &\simeq 
    12\,\mathrm{MeV}
    \times \sqrt{c_\mathrm{b}}  \left( \frac{g_*(T_\mathrm{dec})}{10.75} \right)^{-1/4}
    \sqrt{ \frac{\epsilon m_\phi}{10^{-12}\,\mathrm{eV}} }
    \ .
\end{align}
For the domain wall network to collapse before the BBN, we need $T_\mathrm{dec} \gtrsim 10$\,MeV.
We can also express $a_{\mathrm{min},k}$ as
\begin{align}
    a_{\min,0} = 0
    \ , \quad 
    a_{\min,1} = \pi f_a
    \ .
\end{align}
Then, the {tension of the induced domain wall} is given by 
\begin{align}
    \sigma_a
    \simeq 
    0.237 \pi^2 m_a f_a^2
    \equiv 
    c_\sigma m_a f_a^2
    \ ,
\end{align}
where the coefficient $c_\sigma \simeq 2.34$ corresponds to $\pi^2 \kappa$ for $a_\mathrm{min} = \pi f_a/N_a$.

\subsection{Production of QCD axions}
\label{subsec: production of QCD axions}

Here, we evaluate the QCD axion abundance from the domain wall collapse following Ref.~\cite{Kawasaki:2014sqa}.
The domain walls associated with $\phi$ and $a$ continuously emit $\phi$ and $a$ particles respectively in the scaling regime. However, the most significant particle production occurs during the collapse of these domain walls. 
This is because,
in the scaling regime, the average energy density of the domain walls evolves as $\propto H$, which decreases more slowly than the number density of the emitted particles {$\propto R^{-3}$ with $R$ being the scale factor}. 
We assume that the produced $\phi$ particles immediately decay into the Standard Model particles such as  gluons and photons\footnote{ 
In fact, the $\phi$ particle is coupled to gluons 
since the mixing term $V_{\rm mix}$ comes from the non-perturbative QCD effects. {See \cite{Kitajima:2023cek} for the viable parameter region in this case.}}, and that $a$ is stable and composes dark matter.
Then, the total abundance of $a$ in the later universe is mainly contributed by the emission during the network collapse.
For this reason, we use $m_a = m_a(T_\mathrm{dec})$ to evaluate the axion abundance.

In the scaling regime, the {induced} domain walls emit $a$ particles (and a negligible amount of gravitational waves) at the rate of 
\begin{align}
    \left. 
        \frac{\mathrm{d}\rho_{a\text{-wall}}}{\mathrm{d}t}
    \right|_\mathrm{emit}
    =
    -\frac{\sigma_a}{2 t^2}
    \ ,
\end{align}
where we did not consider the decrease due to cosmic expansion.
If the emitted $a$ has an average energy of 
\begin{align}
    \bar{\omega}_a
    =
    \tilde{\epsilon}_a m_a
    \ ,
\end{align}
where we assume $\tilde{\epsilon}_a$ to be a constant, we obtain the number density of $a$ emitted from the domain walls as
\begin{align}
    n_{a,\mathrm{dec}}(t)
    =
    \frac{\sigma_a}{\tilde{\epsilon}_a m_a t}
    \ ,
\end{align}
for $t < t_\mathrm{dec}$.
Thus, the energy density from the domain wall collapse is given by 
\begin{align}
    \rho_{a,\mathrm{dec}}(t)
    &\simeq 
    \frac{
    \sigma_a}{\tilde{\epsilon}_a t_\mathrm{dec}}
    {\frac{m_{a0}}{m_a}}
    \left( \frac{a(t_\mathrm{dec})}{a(t)} \right)^3
    \ ,
\end{align}
after the emitted $a$ particles become non-relativistic {for $T<T_\mathrm{QCD}$}.
Thus, we can evaluate the ratio of the energy density of $a$ to the entropy density $s$ as
\begin{align}
    \frac{\rho_{a,\mathrm{dec}}}{s}
    &\simeq 
    \frac{2\sigma_a H_\mathrm{dec}}{\tilde{\epsilon}_a}
    \frac{45}{4 \pi^2 g_{*s}(T_\mathrm{dec}) T_\mathrm{dec}^3}
    \nonumber \\
    &\simeq 
    \frac{45}{2 \pi^2 \tilde{\epsilon}_a g_{*s}(T_\mathrm{dec}) }
    c_\sigma m_a(T_\mathrm{dec}) f_a^2
    \sqrt{\frac{\pi^2 g_*(T_\mathrm{dec})}{90}}
    \frac{1}{M_\mathrm{Pl} T_\mathrm{dec}}
    \nonumber \\
    &\simeq 
    \frac{3 \sqrt{10}
    c_\sigma \sqrt{g_*(T_\mathrm{dec})}}{4 \pi \tilde{\epsilon}_a g_{*s}(T_\mathrm{dec}) }
    \frac{m_a(T_\mathrm{dec}) f_a^2}{M_\mathrm{Pl} T_\mathrm{dec}}
    \nonumber \\
    &\simeq 
    0.43\,\mathrm{eV}
    \times \tilde{\epsilon}_a^{-1}
    \left( \frac{g_{*,\mathrm{dec}}}{10.75} \right)^{1/2}
    \left( \frac{g_{*s,\mathrm{dec}}}{10.75} \right)^{-1}
    \left( \frac{f_a}{4 \times 10^9\,\mathrm{GeV}} \right)
    \left( \frac{T_\mathrm{dec}}{12\,\mathrm{MeV}} \right)^{-1}
    \ ,
    \label{eq:QCD axion abundance}
\end{align}
for $T_\mathrm{dec} < T_\mathrm{QCD}$.
If $\rho_{a,\mathrm{dec}}/s = 0.44$\,eV (e.g., see Ref.~\cite{Planck:2018vyg}), the axion emitted from the induced domain wall can explain all dark matter.

We show the parameter region for all dark matter in Fig.~\ref{fig: DM}.
\begin{figure}[!t]
    \begin{center}  
        \includegraphics[width=0.8\textwidth]{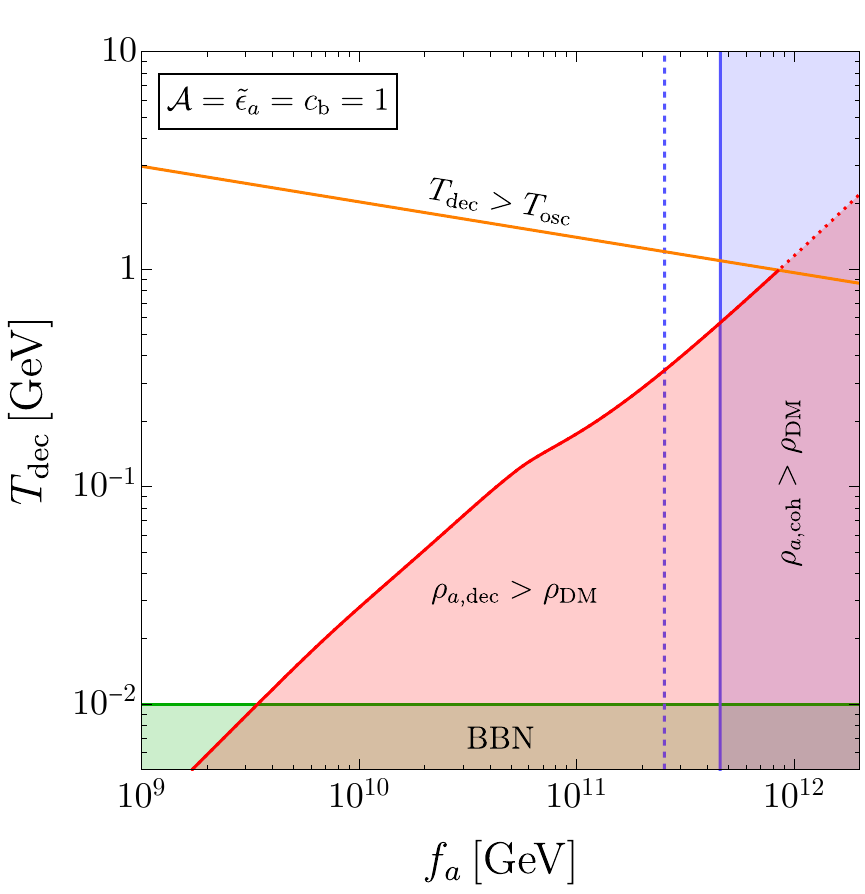}
        \end{center}
    \caption{%
        QCD axion abundance and the constraints on the parameters.
        In the red and blue regions, the QCD axion abundance from the domain wall and coherent oscillations exceeds the observed dark matter abundance, respectively.
        The solid and dashed blue lines correspond to the cases of the minimum and maximum abundance determined by $a_\mathrm{ini}$.
        The green region is excluded by the too-late collapse of the string-wall network.
         {Above the orange line,}
        the string-wall network will collapse before the formation of the domain wall of $a$, and our estimate of the QCD axion abundance (\ref{eq:QCD axion abundance}) becomes invalid.
        Here, we adopted $\mathcal{A} = \tilde{\epsilon}_a = c_\mathrm{b} = 1$ for simplicity, and used the temperature dependence of $g_*$ and $g_{*s}$ given in Ref.~\cite{Saikawa:2018rcs}. 
    }
    \label{fig: DM} 
\end{figure}
The red and blue regions denote the overabundance of the QCD axion contributed by the domain wall collapse and coherent oscillations, respectively.
The solid and dashed blue lines correspond to the minimum and maximum values of $\mathcal{F}(a_\mathrm{ini})$ (see Eq.~\eqref{eq: coherent min & max}).
The green region denotes the limit from the BBN.
{Above the orange line, the string-wall network collapses before the induced domain walls are formed, and thus our estimate of the QCD axion abundance becomes invalid.}
Consequently, we find that the QCD axion can account for all dark matter with $f_a \gtrsim 3 \times 10^9$\,GeV in our scenario.

\subsection{Production of gravitational waves}
\label{subsec: production of GWs}

In the scaling regime, the string-wall network also emits gravitational waves.
Here, we evaluate the spectrum of the gravitational waves following Ref.~\cite{Hiramatsu:2013qaa}.
As in the case of the axion emission, the dominant contribution comes from just before {and during} the network collapse.
Since the frequency of the gravitational waves is determined by the Hubble scale at the emission, the peak frequency of the total gravitational spectrum is estimated by 
\begin{align}
    f_\mathrm{peak,dec}
    \sim
    H_\mathrm{dec}
\end{align}
at $t = t_\mathrm{dec}$.
The peak value of the density parameter of the gravitational waves, $\Omega_\mathrm{GW}$, is estimated as
\begin{align}
    \Omega_\mathrm{GW,dec}^\mathrm{peak}
    =
    \frac{\epsilon_\mathrm{GW} \sigma_\phi^2}{24\pi M_\mathrm{Pl}^4 H_\mathrm{dec}^2}
    \ ,
\end{align}
where $\epsilon_\mathrm{GW}$ is an efficiency parameter of the gravitational wave emission.

Taking into account the redshift after the emission, we obtain the peak frequency at the current time as
\begin{align}
    f_\mathrm{peak,0}
    &=
    \left( \frac{g_{*s,0}}{g_s(T_\mathrm{dec})} \right)^{1/3}
     \frac{T_0}{T_\mathrm{dec}}
    f_\mathrm{peak,dec}
    \nonumber \\
    &\sim
    11\,\mathrm{nHz} \times 
    \left( \frac{g_{*s,\mathrm{dec}}}{10.75} \right)^{-1/3}
    \left( \frac{g_{*,\mathrm{dec}}}{10.75} \right)^{1/2}
    \left( \frac{T_\mathrm{dec}}{100\,\mathrm{MeV}} \right)
    \ ,
\end{align}
and the peak of the current gravitational wave spectrum as
\begin{align}
    \Omega_{\mathrm{GW},0}^\mathrm{peak} h^2
    &=
    \Omega_\mathrm{r,0} h^2 
    \frac{g_{*,\mathrm{dec}}}{g_{*,0}}
    \left( \frac{g_{*s,0}}{g_{*s,\mathrm{dec}}} \right)^{4/3}
    \Omega_\mathrm{GW,dec}^\mathrm{peak}
    \nonumber \\
    &\sim
    2.6 \times 10^{-9} \times 
    \epsilon_\mathrm{GW} \left( \frac{g_{*s,\mathrm{dec}}}{10.75} \right)^{-4/3}
    \left( \frac{T_\mathrm{dec}}{100\,\mathrm{MeV}} \right)^{-4}
    \left( \frac{\sigma_\phi}{2 \times 10^{15}\,\mathrm{GeV}^3} \right)^{-2}
    \ ,
\end{align}
where $T_0 \simeq 2.725$\,K is the current CMB temperature~\cite{Fixsen:2009ug}, $\Omega_{\mathrm{r},0} = 4.15 \times 10^{-5} h^{-2}$ is the current density parameter of radiation~\cite{Planck:2018vyg}, and $h \equiv H_0/(100\,\mathrm{km/s/Mpc})$ is the reduced Hubble constant.
According to numerical simulations, the gravitational wave spectrum around the peak follows $\Omega_{\mathrm{GW},0} \propto f^3$ for $f < f_\mathrm{peak,0}$ and $\propto f^{-1}$ for $f > f_\mathrm{peak,0}$.

We show the current gravitational wave spectrum in Fig.~\ref{fig: GW}.
{The red line denotes the analytical estimate for $\sigma_\phi = 6 \times 10^{15}$\,GeV$^3$ and $T_\mathrm{dec} = 100$\,MeV, while the blue line denotes the numerical result taken from Ref.~\cite{Kitajima:2023cek} (see also Refs.~{\cite{Ferreira:2024eru,Dankovsky:2024zvs}} for similar calculations).
To translate the numerical results to the physical ones,
we used $\sigma_\phi = 1.3 \times 10^{15}$\,GeV$^3$ and $T_\mathrm{dec} = 150$\,MeV.\footnote{In Ref.~\cite{Kitajima:2023cek}, a time-dependent potential bias was used. We adopt their sample with early bias introduction ($\epsilon = 0.025$ in their notation), which approximates a constant bias case.}
}
\begin{figure}[!t]
    \begin{center}  
        \includegraphics[width=0.8\textwidth]{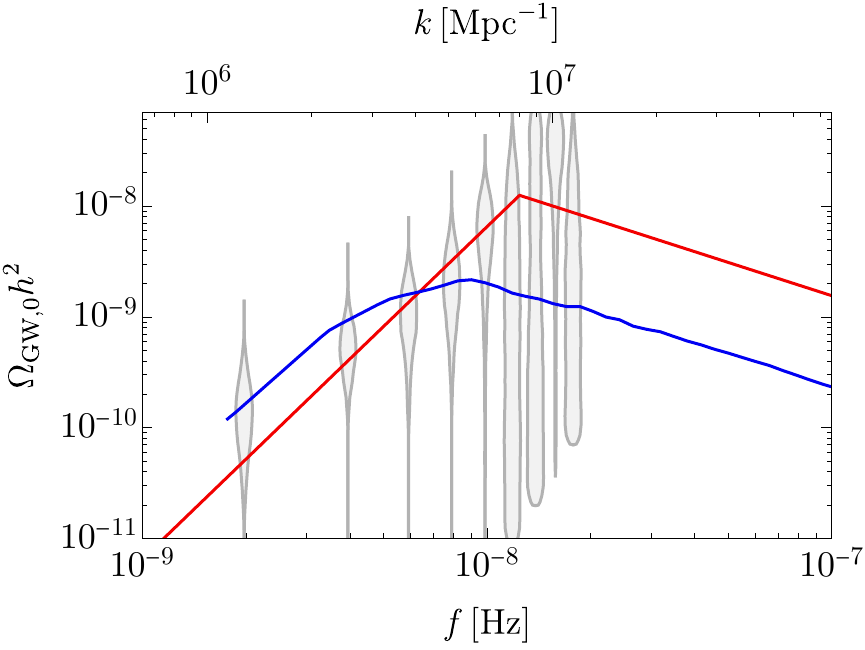}
        \end{center}
    \caption{%
        Gravitational wave spectrum from the domain wall collapse.
        {The red line denotes the analytical estimate with $\sigma_\phi = 6 \times 10^{15}$\,GeV$^3$ and $T_\mathrm{dec} = 100$\,MeV assuming $\mathcal{A} = \tilde{\epsilon}_\mathrm{GW} = 1$.
        The blue line denotes the numerical result for $\sigma_\phi = 1.3 \times 10^{15}$\,GeV$^3$ and $T_\mathrm{dec} = 150$\,MeV translated from Ref.~\cite{Lee:2024oaz}.}
        The gray violins are from the NANOGrav 15\,yr result~\cite{NANOGrav:2023gor}.
        We used the temperature dependence of $g_*$ and $g_{*s}$ given in Ref.~\cite{Saikawa:2018rcs}.
    }
    \label{fig: GW} 
\end{figure}
Although there is uncertainty in the analytical estimate of the gravitational wave spectrum that comes from $\mathcal{A}$ and $\tilde{\epsilon}_a$, $T_\mathrm{dec}$ can be determined if the peak frequency of the gravitational wave spectrum is determined from observations. This allows us to infer the value of $f_a$ by assuming QCD axion dark matter (see Fig.~\ref{fig: DM}). Thus, in principle, $f_a$ can be estimated from observations of gravitational waves.

\section{Summary and discussion}
\label{sec: summary}
{We have demonstrated that heavy axion domain walls can induce domain walls of the QCD axion through the mixing between the heavy axion and the QCD axion, even under the assumption of a pre-inflationary initial condition for the QCD axion. These induced domain walls emerge because the effective $\theta$ parameter varies across the heavy axion domain walls, resulting in a shift in the potential minimum of the QCD axion. 
The induced domain walls overlap with those of the heavy axion and follow the motion of the heavy axion domain walls.
Upon the collapse of the heavy axion domain walls {due to a potential bias}, the induced QCD axion domain walls also collapse. 
{This collapse of the domain walls produces QCD axions, which} can account for {all} dark matter with an axion decay constant as small as ${\cal O}(10^9)$ GeV. 
{Interestingly, despite the assumed pre-inflationary initial condition for the QCD axion, the induced domain walls generate large spatial inhomogeneities, which leads to the formation of axion miniclusters in the later universe.
Furthermore}, this scenario necessitates domain wall collapse near the QCD crossover, which could potentially explain the stochastic gravitational wave background indicated by recent pulsar timing array observations, including those by NANOGrav.
}

In the main text, we focused on pre- and post-inflationary initial conditions for $A$ and $\Phi$, respectively. However, induced domain walls could appear in more general circumstances. For instance, $\phi$ domain walls formed by inflationary fluctuations could also induce QCD axion domain walls. In this scenario, the QCD axion cannot be the dominant component of dark matter due to the superhorizon-scale correlation of induced walls, which would lead to excessive isocurvature perturbations.
Another possibility is that $a$ forms domain walls from inflationary fluctuations, while $\phi$ forms the string-wall network as discussed in the main text. This would result in two types of the $a$ domain walls: the usual one and the induced one with lower tension. Depending on the parameters, the usual $a$ domain walls might appear only in certain domains of $\phi$. Even if the usual QCD axion domain walls disappear, the induced ones persist as long as the $\phi$ walls exist. Since axions produced from induced domain walls do not possess isocurvature perturbations, the overall isocurvature perturbation could be suppressed depending on the parameters.
We have employed various simplifying assumptions regarding mass and tension (see Eqs.~(\ref{eq:hierarchy1}) and (\ref{eq:hierarchy2})) and the form of the potential bias. As previously noted, a detailed study of domain wall evolution in a more general situation is warranted and is left for future work, one of which is given in \cite{Lee:2024toz}.

It is crucial to recognize that our mechanism naturally produces domain walls that are not accompanied by strings. In other words, there is no winding number and the axion is well-defined in the entire space. This feature is important because it allows the formation of domain walls even for string axions (see appendix \ref{app:2}) or axions with extremely large decay constants. Such domain wall formation is typically challenging in these cases, as symmetry restoration is difficult or even impossible.
Analogous to the QCD axion scenario, our induced domain wall mechanism can easily generate axion miniclusters even if the string axion or the axion with a large decay constant initially possesses a nearly homogeneous field value. Moreover, we emphasize that the tension of induced domain walls is generally lower than that of conventional domain walls, since the distance between adjacent minima is determined by the mixing between light and heavy axions. Consequently, for certain values of the mixing parameters, it becomes possible to suppress the abundance of light axions produced by the collapse of induced domain walls.

As an interesting cosmological application, let us consider a light ALP with a pre-inflationary initial condition to explain the recently reported hint of isotropic cosmic birefringence~\cite{Minami:2020odp,Diego-Palazuelos:2022dsq,Eskilt:2022wav,Eskilt:2022cff,Cosmoglobe:2023pgf}.
The cosmic birefringence induced by ALP strings and walls was studied in Refs.\cite{Agrawal:2019lkr,Jain:2021shf,Jain:2022jrp}.
It was pointed out in Refs.\cite{Takahashi:2020tqv,Jain:2022jrp}  that it is difficult to explain the hint for isotropic cosmic birefringence using ALP strings
because it is accompanied by excessive anisotropic birefringence. On the other hand, ALP domain walls without strings naturally induce isotropic birefringence close to the measured value~\cite{Takahashi:2020tqv,Kitajima:2022jzz,Gonzalez:2022mcx,Kitajima:2023kzu}. These domain walls also predict characteristic anisotropic birefringence, which depends on initial fluctuations~\cite{Gonzalez:2022mcx}, while the isotropic component is insensitive to them. 
Domain walls with thermal initial fluctuations (or any fluctuations without significant correlation at superhorizon scales) are vulnerable to population bias, making it non-trivial to have long-lived ALP domain walls without strings from thermal fluctuations. Conversely, ALP domain walls from
inflationary fluctuations remain stable and highly robust against population bias~\cite{Gonzalez:2022mcx}.
Consequently, previous studies focused on inflationary initial fluctuations. However, it is now possible to induce domain walls of a light ALP $a$ using another heavy ALP $\phi$ with post-inflationary initial conditions. Note that the induced domain walls are not attached to any strings of $a$, as we assume pre-inflationary initial conditions for $a$. The stability of the induced walls is ensured by the stability of the string-wall network of $\phi$.
Thus, if only $a$ couples to photons, the expected anisotropic birefringence is similar to that of ALP domain walls with thermal fluctuations~\cite{Takahashi:2020tqv,Kitajima:2022jzz}.
ALP domain walls predict characteristic anisotropic birefringence that depends on initial fluctuations, which can be tested in the near future. 

We also mention that the induced domain wall acts as a force potential sourced by the domain wall~\cite{Kim:2021eye}. If the induced domain wall couples to the standard model fermions, such as electron, nucleon, or muon, then the force causes an effective magnetic field that induces the precisions of the spin towards the direction perpendicular to the domain wall. 
Thus, if the domain walls have sufficiently small tension to avoid the domain wall problem and survive until now, we can probe the existence of induced domain walls. In particular,
by searching for the daily modulation of the effective magnetic field due to the Earth's rotation in certain magnetometers, one can search for the domain wall at a faraway position.

If both light and heavy ALPs couple to photons, they both contribute to cosmic birefringence. The heavier ALP has strings, so the resulting isotropic birefringence is likely to come primarily from the lighter ALP. In this scenario, the isotropic cosmic birefringence comes from the lighter ALP, while the anisotropic cosmic birefringence comes from a combination of domain wall and string effects. The effect on the anisotropic birefringence warrants further numerical studies due to the correlated positions of the two types of domain walls. We expect that this will yield a distinct spectrum of anisotropic cosmic birefringence that may serve as a robust prediction of this scenario.

\section*{Acknowledgments}
This work is supported by JSPS Core-to-Core Program (grant number: JPJSCCA20200002) (F.T.), JSPS KAKENHI Grant Numbers 23KJ0088 (K.M.), 24K17039, (K.M.), 20H01894 (F.T.), 20H05851 (F.T. and W.Y.), 21K20364 (W.Y.), 22K14029 (W.Y.), and 22H01215 (W.Y.), Incentive Research Fund for Young Researchers from Tokyo Metropolitan University (W.Y.), Graduate Program on Physics for the Universe (J.L.) and JST SPRING Grant Number JPMJPS2114 (J.L.). 
This article is based upon work from COST Action COSMIC WISPers CA21106, supported by COST (European Cooperation in Science and Technology).

\appendix

\section{Models for initial conditions} 
\label{app:2}

We have shown a novel phenomenon, where QCD axion domain walls are induced by domain walls of another axion (ALP). We assumed that the initial conditions of the QCD axion and ALP are different: the QCD axion is pre-inflationary and the ALP is post-inflationary.  
In this appendix, we provide natural scenarios that predict the initial conditions assumed in this paper. 

\paragraph{Thermal phase transitions with suitable cosmic temperature}
Let us assume {that} both $\phi$ and $A$ are in {the} broken phase during inflation. 
We notice that the decay constant of $\phi$ should satisfy $f_\phi\ll 10^9\GEV \lesssim f_a$ to ensure prompt decay of $\phi$ particles produced by the collapse of $\phi$ domain walls \cite{Kitajima:2023cek}.
Assuming {that} all the couplings are $\O(1)$ in the UV completion models, and that the maximal temperature of the 
universe, $T_{\rm max}$, satisfies
\begin{align}
f_\phi < T_{\rm max}<f_a,
\end{align} 
we find that the symmetry of $\U(1)_{\rm H}$ {is restored}  due to the thermal potential, while $\U(1)_{\rm PQ}$ remains broken because the radial mode of $A$ is too heavy. 

\paragraph{Phase transitions induced by inflaton}
So far we have used the hierarchy of decay constants and assumed that they remain unchanged during and after inflation. Without relying on this hierarchy, we can still obtain our initial conditions naturally by considering non-minimal interactions:
\begin{align}
{\cal L}\supset -\x_A |A|^2 R -\x_\F |\F|^2 R 
\end{align} 
where $R$ is the Ricci scalar, and $\xi_A$ and $\xi_\Phi$ are dimensionless non-minimal couplings.
During inflation, the PQ fields, $A$ and $\Phi$, acquire the following Hubble-induced masses:
\begin{align}
\d m_A^{2}= 6 \x_A H_{\rm inf}^2,~~ \d m_\F^{2}= 6 \x_\F H_{\rm inf}^2,
\end{align} 
where $H_{\rm inf}$ is the Hubble parameter during inflation.
The couplings are essentially arbitrary but are naturally $\mathcal{O}(1)$, as otherwise, we would need to fine-tune the renormalized couplings between radiative corrections and bare ones.

In the case where
\begin{align}
\xi_A \lesssim-1,\quad \xi_\Phi\gtrsim 1,\quad \text{and}\quad H_{\rm inf}^2\gg f_\phi^2, f_a^2,
\end{align}
$\U(1)_{\rm PQ}$ is broken during inflation, while $\U(1)_{\rm H}$ remains unbroken due to the positive mass squared. Again, we assume all couplings are $\mathcal{O}(1)$, so $f_\phi$ and $f_a$ represent the mass scales of the Higgs bosons (radial modes).

After inflation, the inflaton oscillates around its potential minimum. We obtain an inflaton-dominated universe, which behaves as a matter-dominated universe. The Hubble-induced masses persist as $\sim \xi_{A,\Phi} H^2$, which scales as a scale factor to the power of $-3$.
When the Hubble-induced masses become comparable to the vacuum masses, the latter dominate. Since both symmetries are broken with the vacuum masses, $\U(1)_{\rm H}$ SSB occurs with string network formation, while $\U(1)_{\rm PQ}$ remains broken. The string network follows the scaling solution in the matter-dominated universe. Thus, our assumed initial conditions are naturally realized.
In this scenario, we implicitly assume that thermal effects are negligible {so that the $\U(1)_{\rm PQ}$ is not restored.}

\paragraph{String axion plus field theory ALP}

So far, we have assumed that both axions have field theory UV completion. However, axions also have a nice UV completion in string theory. 
The main discussion may not be compatible with this possibility {unless the mixing is suppressed. This is because} the string QCD axion is likely to have the decay constant higher than $10^{12}\GEV$, which could lead to an overproduction of dark matter. Nevertheless, our model building is useful for having {induced domain walls for generic string axions.} In other words, we would like to {stress} that string axions can also naturally form domain walls {through a mixing with field theoretic axions.}\footnote{
Another simple possibility arises from the inflationary fluctuations of string axions. These fluctuations are typically much smaller than the decay constant of string axions, and domain wall formation, which requires fluctuations to cover the potential hilltop, involves fine-tuning the initial field value for the string axion. However, the string axiverse typically predicts a large number of light axions. If this number is sufficiently large, we may obtain domain walls for string axions from inflationary fluctuations.
}

Let us consider a toy model of the string axion, {where the axion arises from the gauge field in extra dimensions. To be explicit, we consider} {a $\U(1)$ gauge field in $R_{1,3}\times S_1$.} This simplified setup illustrates a string theory UV completion of the axions. The extra dimension is assumed to be a circle $S_1$ with a radius $R$. {We introduce} a $\U(1)$ gauge field  $A_M=(A_\mu, A_5)$ {with} the gauge invariant Lagrangian {in 5D},
\begin{align}
\label{eq:F5}
{\cal L}_5=-\frac{1}{4 g_5^2}F^{MN} F_{MN}, 
\end{align}
where $g_5^2$ is the five-dimensional gauge coupling with the mass dimension $-1$. 
The U(1) gauge transformation  $A_5 \to A_5 + \partial_5 \alpha$ with the boundary condition $\alpha(x,x_5)= \alpha(x, x_5+2\pi R)~ {\rm mod}~ 2\pi$ becomes non-trivial when  $\alpha$ has a winding number, {e.g. $\alpha = x_5/R$}. {We can define the axion field $a(x)$ as
\begin{align}
    \frac{a(x)}{f_a} \equiv \oint dx_5 A_5(x,x_5),
\end{align}
where $f_a$ is the decay constant.
Then, the discrete shift symmetry for the axion $a \to a + 2 \pi f_a$ corresponds to the non-trivial U(1) gauge transformation of $A_5 \to A_5 + \frac{1}{R}$. } 

Considering the zero mode of $A_5$, we can fix the decay constant to obtain the canonically normalized kinetic term for the axion from (\ref{eq:F5}):
\begin{align}
    f_a & =  \frac{1}{\sqrt{2 \pi g_5^2 R}}.
\end{align}
In string theory $R$ is dynamical, but as long as the extra dimension has a finite size we never have the ``symmetric phase", and thus the usual string-wall formation mechanism does not apply. 

\section{Axiverse from non-abelian gauge symmetries}
\label{app:model1}

The effective theory discussed in this paper can be naturally realized if the Peccei-Quinn symmetry is an accidental symmetry arising from an $\SU(N)$ gauge theory, which solves the quality problem of the QCD axion~\cite{DiLuzio:2017tjx,Lee:2018yak, Ardu:2020qmo,Yin:2020dfn}.
In fact, our claim is more generic, and we will show that non-abelian gauge theory can provide an interesting axiverse scenario, where not only the axion mass but also the decay constant spread over wide ranges (cf. $\pi$ axiverse \cite{Alexander:2024nvi}).
In Ref.~\cite{Lee:2018yak}, it was shown that $\SU(N)$ gauge theory generically contains an axion or ALP with good quality due to the spontaneous breaking of the accidental ``baryon number'' symmetry, either through confinement or Higgsing of the $\SU(N)$. This implies that if the fundamental theory contains many non-abelian gauge groups that disappear in the low-energy effective theory, we predict many axions or ALPs.
Here, we further show, using a simple example, that the axiverse naturally provides pions from the confinement of the gauge group mixing with the high-quality axion.
Since they originate from different symmetry breakings, we provide an axiverse with scales of not only mass but also decay constants spread over many orders of magnitude.

As a concrete example, we consider the model given in Ref.~\cite{Yin:2020dfn} (See also Ref.~\cite{Ardu:2020qmo}), where we have a Higgs in a symmetric rank two tensor representation $(N \otimes N)_{\rm sym}$,
\begin{equation}
H: \left( \frac{N^2+N}{2},1 \right) 
\end{equation}
and fermions in the representation
\footnote{We can easily include the $\SU(2)_L$ and $U(1)_Y$ gauge field contribution. } \begin{equation}
Q_L: (N,3), ~~ Q_R: (N,\bar{3}), ~~ \Psi: 6\times (\bar{N},1).
\end{equation}
 Here, the left and right arguments of the parentheses denote the representation under the hidden $\SU(N)$ QCD and the usual $\SU(3)_c$ QCD, respectively. Here the index $L,R$ represents the quarks in fundamental and anti-fundamental representations. 
 
 There is an accidental baryon number symmetry, under which $H$, $Q_{L,R}$, and $\Psi$ have baryon numbers of $2$, $1$,  and $-1$, respecitvely.
 Interestingly, the $\SU(N)\to \SO(N)$ breaking can be obtained by the Higgs mechanism. The Higgs also gives mass to the fermions via the couplings such as $H^* Q_L Q_R$, $H \Psi \Psi$.
At the same time, the accidental baryon number symmetry is spontaneously broken, giving rise to the QCD axion~\cite{Ardu:2020qmo,Yin:2020dfn}.

Even after the Higgsing,
assuming the Higgs-fermion couplings as well as the $SU(3)_c$ coupling are sufficiently small,
the theory still has an approximate $\SU(12)$ symmetry, which should be further spontaneously broken due to the strong dynamics of $\SO(N).$ 
For certain Yukawa couplings of the Higgs field, the $\SU(3)_c$ gauge symmetry should not be spontaneously broken thanks to the mass persistency condition. On the other hand, the $\SU(12)$ includes the symmetry that is anomalous to the $\SU(3)_c$ subgroup. 
If $\SU(12)$ is sufficiently spontaneously broken, we have a pion coupled to the Chern-Simons term of the $\SU(3)_c$ similar to the $\pi^0$ coupled to the QED in the Standard Model.

In the above setup, the QCD axion arises from the spontaneous breaking of the baryon number symmetry which has a good quality. On the other hand,
a neutral pion can couple to the $\SU(3)_c$ Chern-Simons term, and so it behaves as an ALP that is mixed with the QCD axion. The ALP has a decay constant around the confinement scale, but the QCD axion has a decay constant around the Higgs VEV scale. 
Although the details of the chiral symmetry phase transition are non-trivial and require  further investigation, we have shown that the Lagrangian discussed in the main part can naturally appear as the low energy effective theory from a UV completion that solves the quality problem of the QCD axion.

This is also a concrete example that if the fundamental theory contains many non-abelian gauge theories, we expect to have a much richer ALP and axion spectra and decay constants, giving rise to a new type of axiverse. 

\section{Properties of induced domain walls}
\label{app: DW tension}

In this appendix, we discuss the configuration and tension of the induced domain wall.

First, we consider $|N_a a_\mathrm{min}/f_a| \ll 1$.
In this case, the equation of motion can be approximated by 
\begin{align}
    \partial_z^2 \bar{a}(z) 
    &\simeq
    m_a^2 \bar{a}(z)
    &(z < 0) 
    \ ,
    \\
    \partial_z^2 \bar{a}(z)
    &\simeq
    m_a^2 ( \bar{a}(z) - a_\mathrm{min} )
    &(z > 0)
    \ ,
\end{align}
and we obtain the solution as
\begin{equation}
    \bar{a}(z) 
    \simeq 
    \left\{
    \begin{array}{cc}
        \frac{e^{m_a z}}{2} a_\mathrm{min}
        & (z < 0)
        \\
        \left( 1 - \frac{e^{-m_a z}}{2} \right) a_\mathrm{min}
        & (z > 0)
    \end{array}
    \right.
    \ .
\end{equation}
From this configuration, we obtain the tension of the induced domain wall as  
\begin{align}
    \sigma_a 
    \simeq
    \frac{m_a a_\mathrm{min}^2}{4}
    \ .
\end{align}
The $z$-dependences of the axion fields and $\rho_a$ in this case are shown in Fig.~\ref{fig: DW configuration}.

Next, we consider $N_a a_\mathrm{min}/f_a \sim 2 \pi$.
Then, the equation of motion can be approximated by 
\begin{align}
    \partial_z^2 \bar{a}(z) 
    &\simeq
    \frac{m_a^2 f_a}{N_a} 
    \sin \left(N_a \frac{\bar{a}(z)}{f_a} \right)
\end{align}
for all $z$.
This is the same as in the standard discussion on domain walls with a single axion.
The solution is given by 
\begin{equation}
    \bar{a}(z) 
    \simeq 
    \frac{4 f_a}{N_a} \arctan \left[ e^{m_a z} \right]
    \simeq 
    \frac{2}{\pi} a_\mathrm{min}
    \arctan \left[ e^{m_a z} \right]
    \ .
\end{equation}
Then, the tension of the induced domain wall becomes
\begin{align}
    \sigma_{a,\mathrm{max}} 
    =
    \frac{8 m_a f_a^2}{N_a^2}
    =
    \frac{2 m_a a_\mathrm{min}^2}{\pi^2}
    \ .
\end{align}

Finally, we show the domain wall configuration for some values of $N_a a_\mathrm{min}/f_a$ in Fig.~\ref{fig: abar}.
While the configurations have a similar feature, the detailed shape depends on $N_a a_\mathrm{min}/f_a$.
\begin{figure}[!t]
    \begin{center}  
        \includegraphics[width=0.8\textwidth]{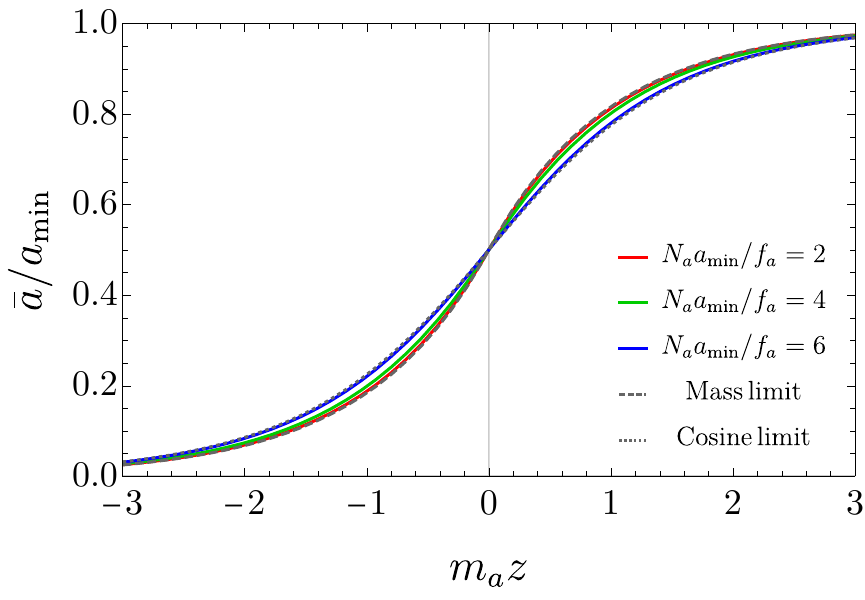}
        \end{center}
    \caption{%
        Spatial configuration of the induced domain walls, $\bar{a}/a_\mathrm{min}$.
        The colored lines denote different values of $a_\mathrm{min}$, and the dashed and dotted lines denote the limit of $|N_a a_\mathrm{min}/f_a| \ll 1$ and $|N_a a_\mathrm{min}/f_a| \to 2\pi$.
    }
    \label{fig: abar} 
\end{figure}

\bibliographystyle{apsrev4-1}
\bibliography{ref}

\end{document}